\documentclass[preprint,12pt,graphics]{elsarticle}

\usepackage{amssymb}

\usepackage{lineno}

\journal{Astroparticle Physics}
\begin{document}

\begin{frontmatter}

\title{Analysis of the Data from Compton X-ray Polarimeters which 
Measure the Azimuthal and Polar Scattering Angles}


\author[a1]{H. Krawczynski}
\ead{krawcz@wuphys.wustl.edu, Tel. 314 935 8553, Fax. 314 935 6219}
\address[a1]{Washington University in St. Louis,
   Department of Physics and the McDonnell Center for the Space Sciences,
   1 Brookings Dr., CB 1105,
   St Louis,
   MO 63130}

\begin{abstract}
X-ray polarimetry has the potential to make key-contributions to our 
understanding of galactic compact objects like binary black hole systems 
and neutron stars, and extragalactic objects like active galactic nuclei,
blazars, and Gamma Ray Bursts. Furthermore, several particle astrophysics
topics can be addressed including uniquely sensitive tests of Lorentz
invariance. In the energy range from 10 keV to several MeV,
Compton polarimeters achieve the best performance.
In this paper we evaluate the benefit that comes from using the azimuthal 
and polar angles of the Compton scattered photons in the analysis, 
rather than using the azimuthal scattering angles alone. 
We study the case of an ideal Compton polarimeter and show that a Maximum Likelihood 
analysis which uses the two scattering angles lowers the Minimum Detectable Polarization 
(MDP) by $\approx$20\% compared to a standard analysis based on the azimuthal scattering angles alone.
The accuracies with which the polarization fraction and the polarization direction can be measured
improve by a similar amount. We conclude by discussing potential applications of Maximum 
Likelihood analysis methods for various polarimeter experiments.   
\end{abstract}
\end{frontmatter}
\section{Introduction}
\label{intro}
Some of the most interesting astrophysical objects are too small to be spatially resolved by 
present-day telescopes: the Schwarzschild radius of a galactic stellar mass black hole 
of $M_1\,=$ 10 $M_{\odot}$ solar masses at a distance of $d_3$ kpc corresponds to 
an angular extent of 2$\times 10^{-4}$ $M_1/d_3$ $\mu$arcsec; 
for an extragalactic supermassive black hole of $M_9\,=$ $10^{9}$ $M_{\odot}$ solar masses 
at a distance of 20 $d_6$ Mpc, the Schwarzschild radius corresponds to an
angular extent of $M_9 \,/\, d_6$ $\mu$arcsec; 
a time of $\Delta t\,=$ 1000 sec (in the observer frame) after the explosion 
of a Gamma Ray Bursts (GRB) at redshift $z\,=$ 1, a GRB jet has an angular extent 
of $1.2 \times 10^{-10}$ $(\Gamma/10^3)^2$ $(\Delta t/1000 \,\rm sec)$  
$(\sin{(\theta)}/\sin{(1^{\circ})})$ 
$\mu$arcsec if we assume (somewhat unrealistically) a constant jet expansion velocity with a 
bulk Lorentz factor $\Gamma$ along an angle $\theta$ to the line of sight.
The numbers demonstrate the difficulties to image these objects. In the absence of 
imaging information, X-ray spectropolarimetric observations can deliver the additional
information required to decide between models which cannot be distinguished 
based on spectroscopic data alone \citep{Rees:75,Shap:75,Shap:76,Mesz:88,Rees:94,Lei:97,Weis:06,Bell:10,Kraw:10}. 
An X-ray polarimeter can deliver for each time interval not only an energy 
spectrum of the X-ray flux $F(E)$, but also the energy spectra of the X-ray polarization degree $a_0(E)$ 
and the X-ray polarization direction $\phi_0(E)$. 
%
X-ray polarimetry will make a big leap when NASA will launch the Gravity and Extreme Magnetism SMEX ({\it GEMS}) mission in 2014 \citep{Swan:10}.
{\it GEMS} will use two Wolter type mirror assemblies to focus 2-10 keV X-rays onto two 
photo-effect polarimeters. Each polarimeter is made of a photoelectron-tracking time 
projection chamber. A student experiment will give additional polarization sensitivity at 0.5 keV.

In this paper, we discuss the analysis of data from polarimeters in the $>10$ keV to several MeV 
energy range based on the detection of Compton-scattered photons. 
Compton polarimeters use the fact that photons scatter preferentially  
perpendicular to the orientation of the electric field vector of a
polarized X-ray beam. Several past and present satellites were equipped 
with detector configurations which -- although not designed for Compton polarimetry -- 
were used to constrain the polarization of the X-ray emission from cosmic 
sources based on the analysis of Compton events (see \citep{Lei:97,Bell:10} and references therein).
Recently, two experiments on board of the {\it INTEGRAL} (International Gamma-Ray Astrophysics Laboratory) 
satellite were used to measure the polarization
of the 100 keV to 1 MeV emission from the Crab Pulsar and the Crab Nebula.
Events scattered between different Ge detectors of the {\it SPI} (SPectrometer on {\it INTEGRAL}) 
detector were used to measure a polarization degree of 46\%$\pm$10\% and a polarization direction
aligned with the orientation of the X-ray jet \citep{2008Sci...321.1183D}.
The analysis of events scattered between different CdTe detectors of the {\it IBIS} 
(Imager on-Board {\it INTEGRAL} Satellite) detector confirmed the large polarization fraction
of the hard X-rays \citep{Foro:08}. In a similar fashion, Compton polarimetry was used
with the Ge detectors onboard of the RHESSI experiment to constrain the 
polarization of solar flares in the $>$200 keV energy range \citep{Bogg:06,Suar:06}.
A report of polarized $\gamma$-ray emission from a 
Gamma-Ray Bursts remained controversial after several independent analyses did not confirm 
the result \citep{Cobu:03,Rutl:04,Wigg:04}.
The soft gamma-ray telescope on board of the Japanese/US ASTRO-H mission 
(launch foreseen in 2013) will be able to measure the polarization of hard X-rays
by detecting Compton scattered photons with a combination of Si pad detectors 
and Cadmium Telluride (CdTe) pixel detectors \citep{Taka:08}.      

A considerable number of dedicated Compton X-ray polarimeters have been described in the literature, see
for example \citep{Cost:95,Lei:97,Mizu:05,Kraw:09,McCo:09a,McCo:09b,Grei:09,Bell:10,Vada:10} and references therein.
Whereas some polarimeters are designed to measure exclusively the azimuthal scattering angles
(e.g.\ \citep{Mizu:05,McCo:09a}), other experiments also give information about the polar 
scattering angle on event to event basis (e.g.\ \citep{Taka:08,Kraw:09,Grei:09}). 
The analyses used so far for both types of experiments are based on fitting a constant offset
plus a sine curve to the azimuthal scattering angle distributions. 
In this paper, we explore for the first time the sensitivity gain that can be achieved when 
using both, the azimuthal and the polar scattering angles in the analysis with the help of a
Maximum Likelihood analysis method. In Sects.\ \ref{basics} and \ref{methods} we motivate 
the use of the polar and the azimuthal scattering angles based on the properties of the 
Klein-Nishina cross section and describe the standard analysis method and the 
Maximum Likelihood analysis method. In Sect.\ \ref{perf} we discuss simulations 
of an ideal Compton polarimeter and evaluate the sensitivity gain achieved with 
the Maximum Likelihood analysis method. In Sect.\ \ref{discussion} 
we summarize the results.
\section{Compton Scatterings: Basics}
\label{basics}
Compton scatterings are governed by Compton's formula and by the Klein-Nishina cross section.
Compton's formula reads:
\begin{equation}
\Delta \lambda = \frac{h}{m_e\, c}\,\left(1-{\rm cos}\theta \right)
\end{equation}
where $h$ is Planck's constant, $m_{\rm e}$ is the electron mass, 
$c$ is the speed of light, and $\theta$ the polar scattering angle.
The Klein-Nishina cross section is given by (see e.g.\ \citep{Eva:55}):
\begin{equation}
\frac{d\sigma}{d\Omega}\,=\,
\frac{r_0^2}{2}
\frac{k_1^2}{k_0^2}
\left[
\frac{k_0}{k_1}+\frac{k_1}{k_0}-2 {\rm sin}^2 \theta {\rm cos}^2 \eta
\right]
\end{equation}
with $r_0$ the classical electron radius, {\bf k}$_0$ and {\bf k}$_1$ the
wave-vectors before and after scattering, $\theta$ the scattering angle
(the angle between {\bf k}$_0$ and {\bf k}$_1$), and $\eta$
the azimuthal scattering angle, i.e.\ the angle between the 
electric vector of the incident photon and the scattering plane. 
Averaging over the azimuthal scattering angles gives the cross 
section for a scattering of a polarized photon with a polar scattering angle $\theta$
irrespective of $\eta$: 
\begin{equation}
\frac{d\sigma}{d\Omega}(\theta)\,=\,
\frac{1}{2}\left(\frac{d\sigma}{d\Omega}\right)_{\eta=0}
+\frac{1}{2}\left(\frac{d\sigma}{d\Omega}\right)_{\eta=\pi/2}
\,=\,
\frac{r_0^2}{2}
\frac{k_1^2}{k_0^2}
\left[
\frac{k_0}{k_1}+\frac{k_1}{k_0}-{\rm sin}^2 \theta 
\right]
\label{knt}
\end{equation}
This cross section equals the Klein-Nishina cross section
for unpolarized incident photons.

As linearly polarized X-rays are preferentially scattered perpendicular to the orientation
of the electric field vector, the probability density function of the azimuthal 
scattering angle $\eta$ exhibits a modulation of the form:
\begin{equation}
P(\eta)\,=\, \frac{1}{\pi} \left(1+b \cos{\left[2\left(\eta-\frac{\pi}{2}\right)\right]}\right)
\label{eta}
\end{equation}
The modulation amplitude $b$ of the azimuthal scattering angle distribution 
can be computed from the Klein-Nishina cross section:
\begin{equation}
b(\theta) \,=
\frac{
\left(\frac{d\sigma}{d\Omega}\right)_{\eta=\pi/2}-\left(\frac{d\sigma}{d\Omega}\right)_{\eta=0}}
{\left(\frac{d\sigma}{d\Omega}\right)_{\eta=\pi/2}+\left(\frac{d\sigma}{d\Omega}\right)_{\eta=0}}
\,=\,\frac{{\rm sin^2}\theta}{k_0/k_1+k_1/k_0-{\rm sin}^2\theta}
\label{b}
\end{equation}
\begin{figure}[tb]
\begin{center}
\includegraphics[width=8.8cm]{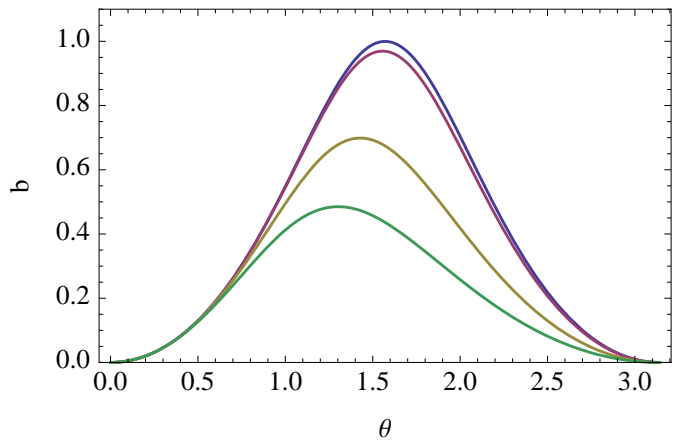}
\caption{\label{muFig} The modulation amplitude $b$ of the azimuthal scattering angle 
from Compton scatterings as function of the polar scattering angle for photon 
energies of 10 keV, 100 keV, 500 keV, and 1 MeV (from top to bottom).}
\end{center}
\end{figure}
Fig.\ \ref{muFig} shows the modulation amplitude as function of $\theta$ for several 
initial photon energies. The panel shows that events with scattering angles close to $\pi$/2
exhibit a larger modulation amplitude. As a consequence, those events have a higher information content
than events with scattering angles close to 0 and close to $\pi$.     
\section{Analysis Methods}
\label{methods}
In the standard analysis the azimuthal scattering angles $\phi$ are histogrammed. The polarization
is determined by fitting the model
\begin{equation}
n(a_0,\phi_0;\phi)\,=\,\bar{n} \left(1+a_0 \, \mu \cos{\left[2(\phi-\phi_0)\right]}\right)
\end{equation}
with a $\chi^2$-fit to the histogram. The value $\bar{n}$ is the mean number of entries 
in each bin of the histogram: $\bar{n}\,=$ $n_0/N$, if $n_0$ events were recorded and 
the histogram has $N$ bins. 
The fitting parameter $a_0$ is the fractional polarization of the beam and equals 1 for
a 100\% polarized beam and 0 for an unpolarized beam.
The parameter $\mu$ is the ``modulation factor'' which gives the modulation 
amplitude for a 100\% polarized beam. In general, $\mu$ depends on the
physics of Compton scatterings and on the design of the polarimeter. 
In the following we will consider an ``ideal polarimeter'' which detects
all Compton scattered photons and gives the scattering angles with 
very high precision. In this particular case, $\mu$ is simply the modulation 
amplitude $b$ averaged over all solid angles and the Klein-Nishina cross section. 
For primary photon energies of 10 keV, 100 keV, and 1 MeV the averaged
modulation factor equals 0.5, 0.48, and 0.25, respectively.
The angle $\phi_0$ is the fitted polarization direction (in the coordinates 
of the polarimeter).

If $a_{\rm T}$ and $\phi_{\rm T}$ are the true polarization fraction and polarization direction,
the probability distribution of the measured values $a_0$ and $\phi_0$ is given by 
\citep{Vino:65,Jenk:68,Weis:06}:
\begin{equation}
P(a_0,\phi_0)\,=\,
\frac {N\, \bar{n}^2 a_0 \mu^2} {4\pi \sigma^2}
\exp{\left[
-\frac{N\, \bar{n}^2 \mu^2}{4 \sigma^2}
\left(a_0^{\,2}+a_{\rm T}^{\,2}- 2 a_0 a_{\rm T} \cos{(\phi_0-\phi_{\rm T})}
\right)
\right]} 
\end{equation}
In the absence of background, the statistical error on the number of counts per bin is given
by Poisson statistics: $\sigma\,=$ $\sqrt{\bar{n}}$.
Integration of this distribution over $\phi_0$ gives the probability to measure a polarization 
degree $a_0$ independent of $\phi_0$:
\begin{equation}
P(a_0)\,=\,
\frac{N\, \bar{n}^2 a_0 \mu^2}{2 \sigma^2}
\exp{\left[
-\frac{N\, \bar{n}^2 \mu^2}{4 \sigma^2}
\left(a_0^{\,2}+a_{\rm T}^{\,2}\right)\right] I_0\left(\frac{N\, \bar{n}^2 \mu^2 a_0 a_{\rm T}}
{2 \sigma^2} \right)}
\end{equation}
with $I_0$ the modified Bessel function of order zero.
This probability distribution can be used to determine the minimum fractional polarization
which can be measured at a certain confidence level. The Minimum Detectable Polarization (MDP) 
is defined as the fractional polarization $a_{1\%}$ which will be measured 
in the absence of any true polarization $a_{\rm T}\,=$ 0 with a chance probability of 1\%:
\begin{equation}
P(a>a_{1\%})\,=\,\int_{a_{1\%}}^{\infty}P(a_0)\,da_0\,=\,1\%
\end{equation}
Integration gives for $\sigma\,=$ $\sqrt{\bar{n}}$:
\begin{equation}
MDP\,\approx \frac{4.29}{\mu\,\sqrt{n_0}}
\label{MDP}
\end{equation}

As an alternative to the standard method, we evaluate below a Maximum Likelihood analysis.
The method uses an event list as input:
\begin{equation}
{\cal{L}}\,=\,\left\{(E_i,\theta_i,\phi_i),i=1 ... n_0 \right\}
\end{equation}
with $E_i$, $\theta_i$, and $\phi_i$ being the energy, polar scattering angle, and azimuthal
scattering angle of the i$^{\rm th}$ event, assumed here to be measured with high precision.
The polarization degree $a_0$ and polarization direction $\phi_0$ are determined by maximizing 
the Likelihood function:
\begin{equation}
\lambda\left(a_0,\phi_0\right)\,=\,
\prod_{i=0}^{n_0}\,
P\left(\phi_i;a_0,\phi_0,E_i,\theta_i\right)
\label{M1}
\end{equation}
In practice, it is more convenient to maximize the logarithm of the 
likelihood function $\ln{\lambda}$ than to maximize $\lambda$. 
The probability of measuring the azimuthal scattering angle $\phi_i$ given
the $a_0$, $\phi_0$, $E_i$, and $\theta_i$ is given by the expression:
\begin{equation}
P\left(\phi_i;a_0,\phi_0,E_i,\theta_i\right)\,=\, 
\frac{1}{\pi} \left(1+a_0 \mu(E_i,\theta_i) \cos{\left[2(\phi_i-\phi_0)\right]}\right)
\label{M2}
\end{equation}
with $\mu(E_i,\theta_i)$ being the modulation factor for the incident photon energy $E_i$ and 
the polar scattering angle $\theta_i$. For the ideal polarimeter, $\mu$ is given 
by Equ.\ (\ref{b}). The maximization of $\ln{\lambda}$ gives the maximum likelihood
estimates $a_{\rm 0,ML}$ and $\phi_{\rm 0, ML}$ of the polarization fraction and
polarization direction, respectively. 
  

We recommend to calculate the statistical significance of the detection 
of a polarized signal by simulating the background and the signal
assuming the null hypothesis (an unpolarized signal) and calculating 
the fraction of simulated event lists with best-fit polarization 
degrees exceeding the observed one.

One can estimate the uncertainty on the fit-parameters by mapping
the Bayesian posterior probability density function \citep{Naka:10} (see also \citep{Weis:10}):
\begin{equation}
p\left(a_0,\phi_0\right)\,=\,\frac{\lambda(a_0,\phi_0)}
{\int\limits_0^1 da_0' \int\limits_0^{\pi} d\phi_0'\,\lambda(a_0',\phi_0')}
\end{equation}
The function $p$ can be used to plot a region $\cal{R}_{\alpha}(\beta)$ (or several regions) 
which contains the true value $a_{\rm T}$ and $\phi_{\rm T}$ with a certain 
probability $\alpha$: 
\begin{equation}
{\cal{R}}_{\alpha}(\beta)\,=\,\left\{(a_0,\phi_0)\left|\right. 
p\left(a_0,\phi_0\right)>\beta,a_0\in\left[0,1\right], \phi_0\in\left[0,\pi\right]\right\}
\end{equation}
with $\beta$ is chosen such that
\begin{equation}
\int\limits_{\cal{R}_{\alpha}(\beta)} da_0'\, d\phi_0' \,\, p(a_0',\phi_0')
\,=\,
\alpha
\end{equation}
\section{Achieved Performance}
\label{perf}
\begin{table}[t]
\begin{tabular}{p{2.8cm}|p{2.8cm}p{2.8cm}p{2.8cm}}
\hline
$n_0$ & MDP Equ.\ (\ref{MDP}) $\left[\%\right]$& MDP Stand.\ Method\ $\left[\%\right]$& MDP Max.\ Lik.\ Method $\left[\%\right]$\\ \hline\hline
1000 &28.3 & 28.3 & 23.8\\
3000 &16.3 & 16.2 & 13.7\\
10000 &8.9 & 9.0 & 7.4\\
50000 &4.0 & 3.9 & 3.3\\ 
\hline \hline
\end{tabular}
\caption{\label{t1} Summary of simulation of data sets for a primary photon energy of 100 keV.
For each number of Compton events, 10,000 data sets were simulated assuming no inherent polarization
$a_{\rm T}\,=$ 0. For each data set the polarization fraction $a_0$ was reconstructed with 
various methods. The $a_0$-distributions were used to determine the Minimum Detectable 
Polarization (MDP) being the value that is larger than 99\% of the reconstructed MDP values.  
}
\end{table}
We evaluated the performance of the standard analysis and the Maximum Likelihood analysis 
by generating simulated data sets for $a_{\rm T}\,=0$ and the $n_0$-values
given in Table \ref{t1}. For each parameter configuration, 10,000 data sets were simulated.
We limited the study to primary photon energies of 100 keV. 
Individual events were generated by first drawing a polar scattering angle 
from the distribution given by Equ.\ (\ref{knt}). Subsequently, an azimuthal scattering angle 
was generated according to the distribution given by Equ.\ (\ref{eta}). The event lists 
were analyzed with the two analysis methods described above. 

\begin{figure}[tb]
\begin{center}
\includegraphics[width=10cm]{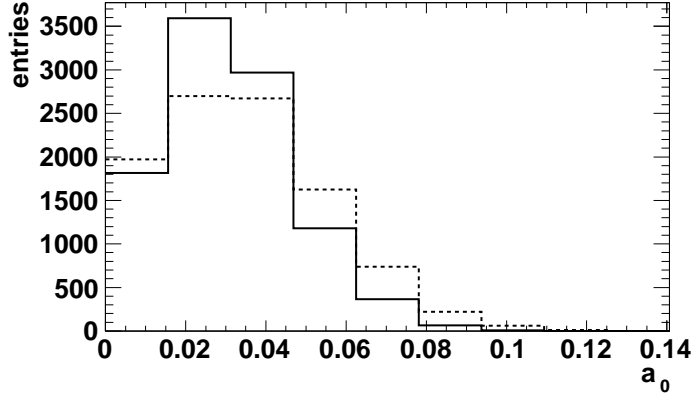}
\caption{\label{a0} Distribution of the reconstructed polarization fraction $a_0$
for a true polarization fraction of $a_{\rm T}\,=$ 0. The solid lines show the results 
for the Maximum Likelihood analysis which uses the polar and azimuthal scattering angles 
of the Compton events; the dashed lines show the results for the standard analysis 
which uses only the azimuthal scattering angles ($n_0\,=$ 10,000).}
\end{center}
\end{figure}
Figure \ref{a0} shows the distributions of the best-fit polarization fraction $a_0$ for
the case of no polarization $a_{\rm T}\,=$ 0 for data sets with $n_0\,=$ 10,000 events. 
The Maximum Likelihood method gives on average smaller best-fit polarization values.
The MDP is calculated by determining the $a_0$-value 
which is larger than 99\% of all fitted values. For $n_0\,=$ 10,000 the 
standard method and the Maximum Likelihood method give MDPs of 0.089 and 0.074, respectively. 
Thus, the Maximum Likelihood method results in a reduction of the MDP by 21\%.
It should be noted that the observed MDP for the standard method agrees 
well with the estimate from Equ.\ (\ref{MDP}).
\begin{figure}[tb]
\begin{center}
\includegraphics[width=10cm]{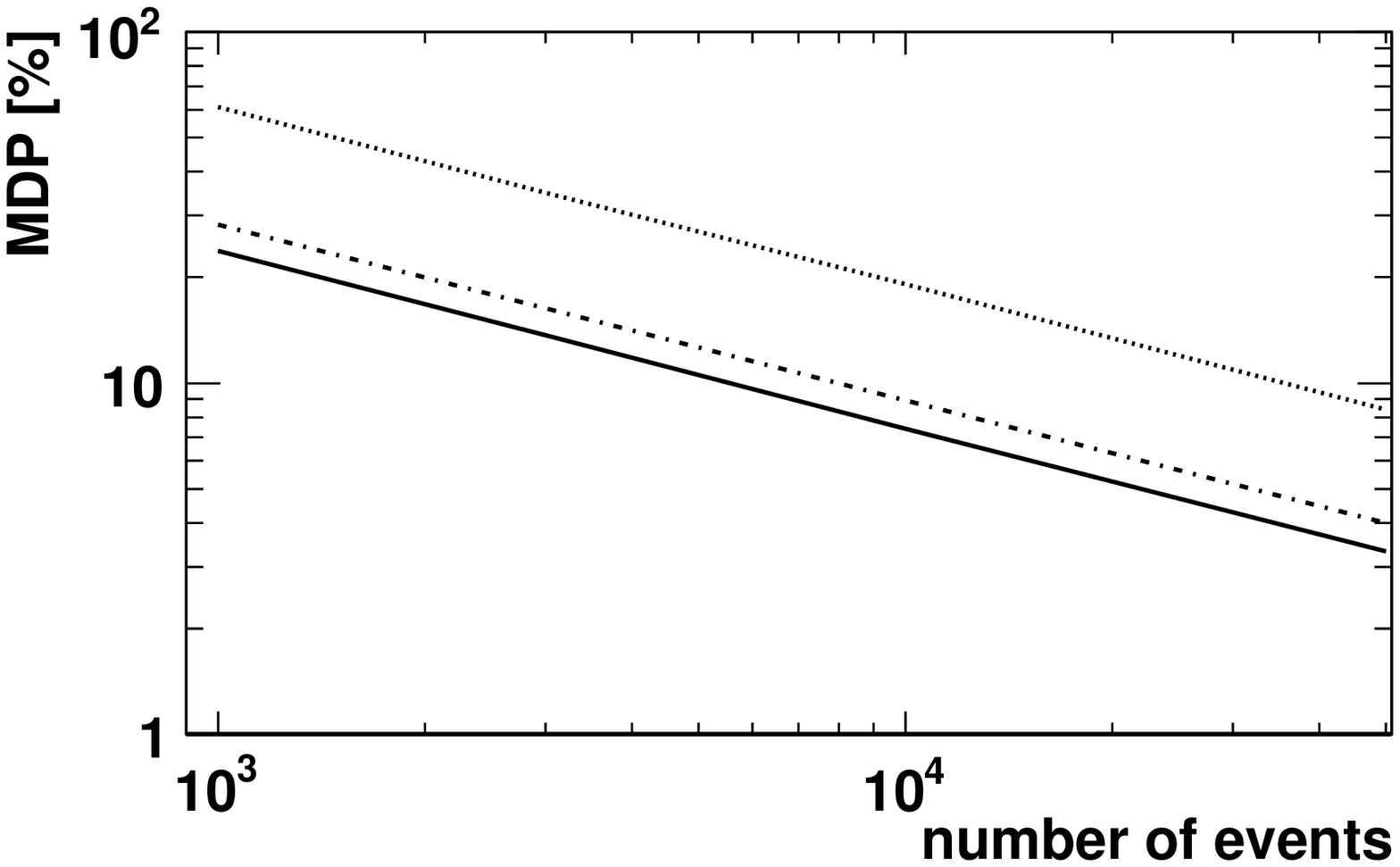}
\caption{\label{a1} Minimum Detectable Polarization (MDP) as function of the number of
Compton events for (from below to above) the Maximum Likelihood method (solid line),
the standard method (dashed line), and for a simpler method (dotted line) in which
the polarization degree is derived from the maximum and minimum number 
of entries in the azimuthal scattering angle histogram. The Maximum Likelihood 
method leads to an improvement by 21\% compared to the standard method.}
\end{center}
\end{figure}

Figure \ref{a1} shows the MDPs achieved with the two methods for $n_0$-values 
ranging from 1000 to 50,000 for the Maximum Likelihood method, the standard method,
and a simpler method. In the case of the ``simple method'' the polarization degree is 
derived from the maximum number $n_{\rm max}$ and minimum number 
$n_{\rm min}$ of entries in the azimuthal scattering angle 
histogram according to
\begin{equation}
a_0\,=\,\frac{1}{\mu} \, \frac{n_{\rm max}-n_{\rm min}}{n_{\rm max}+n_{\rm min}}.
\end{equation}
Whereas the standard method uses the contents of all bins in the histogram to 
estimate $a_0$, the simple method uses only the contents of two bins (the bins
with the minimum and maximum number of counts). It is thus expected that standard
method gives more precise estimates of $a_0$. In the absence of a true polarization
signal, the standard method should thus give on average smaller $a_0$-values than 
the simple method. As a result, the standard method leads to a smaller $a_0$-value
that can be detected at a certain confidence level than the simple method.
Figure \ref{a1} indeed shows that the simple method performs markedly poorer 
than the standard method. Whereas the MDP achieved with the standard method 
is independent of the number of bins used in the azimuthal scattering histogram
(as long as the number of bins is not too small to smear out the modulation, and not
too large to reduce the number of entries per bin to values below 10),
the simple method performs poorer the more bins are used in this histogram.

Compared to the simple method and the standard method, the Maximum Likelihood leads to a 
reduction of the MDP by 255\% and 21\%, respectively, independent of $n_0$.
As mentioned before, the improvement relative to the simple method depends on
the number of bins used in the azimuthal scattering angle histogram.
\begin{figure}[tb]
\begin{center}
\hspace*{-1cm}
\includegraphics[width=15cm]{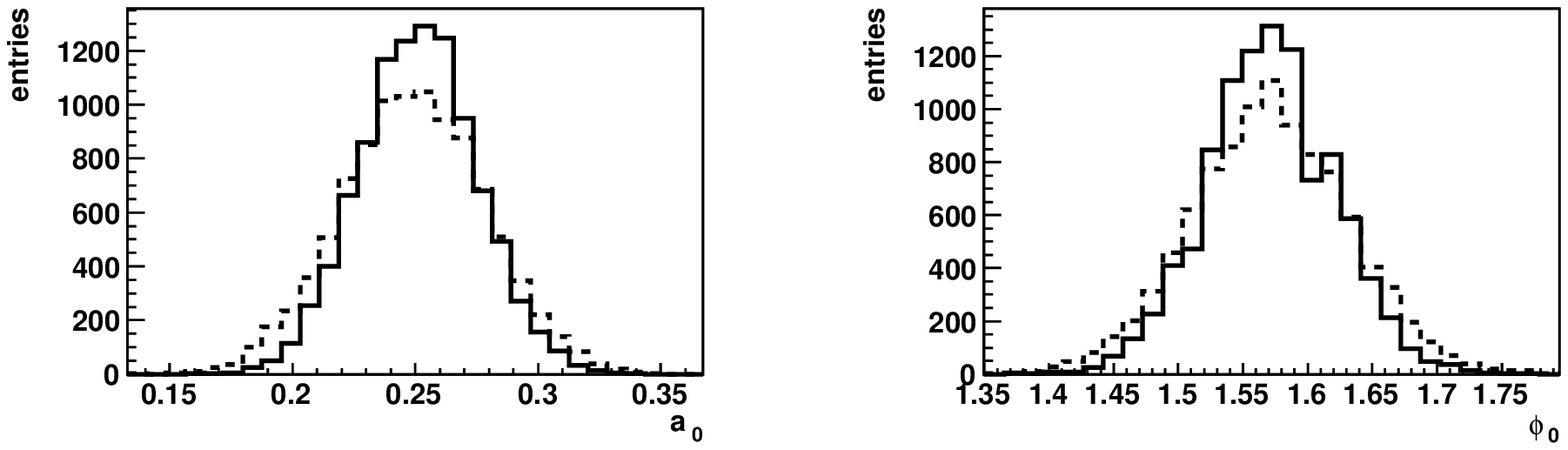}
\caption{\label{a2} Distribution of the reconstructed polarization fraction $a_0$ (left panel)
and the reconstructed polarization degree $\phi_0$ (right panel) for a true polarization fraction
of $a_{\rm T}\,=$ 0.25 and a true polarization direction $\phi_{\rm T}\,=$ $\pi/2$. The solid lines show the results for
the Maximum Likelihood analysis which uses the polar and azimuthal scattering angles of the Compton
events; the dashed lines show the results for the standard analysis which uses only the 
azimuthal scattering angles (both panels are for $n_0\,=10,000$).}
\end{center}
\end{figure}
Figure \ref{a2} shows the distributions of the fitted $a_0$ and $\phi_0$-values
for $a_{\rm T}\,=$0.25, $\phi_{\rm T}\,=$ $\pi/2$, and $n_0\,=$ 10,000.
The standard analysis and the Maximum Likelihood analysis give 68\% confidence intervals 
on $a_0$ of 29\% and 24\%, respectively. For the polarization angle $\phi_0$ the   
68\% confidence intervals are 0.06 rad (3.4$^{\circ}$) and 0.048 rad (2.7$^{\circ}$) 
for the two respective methods.
\section{Summary and Discussion}
\label{discussion}
Photons of a linearly polarized X-ray beam Compton scatter preferentially into the direction 
perpendicular to the preferred orientation of the electric field vector.
The modulation of the azimuthal scattering angle distribution depends on the polar scattering angle
$\theta$. The modulation is most pronounced for photons with polar scattering angles close 
to $\pi/2$ (slightly less for higher energies), and weakest for forward-scattered 
and back-scattered photons. In this paper, we have described a Maximum Likelihood 
analysis which uses the polar and azimuthal scattering angles of all detected Compton 
events to estimate the polarization degree and the polarization direction.
For an ideal Compton polarimeter, the Maximum Likelihood analysis results in a reduction 
of the MDP by 21\%.  The accuracies ($\sigma_{68\%}$-values) with which non-zero 
polarization degrees and polarization directions can be measured are reduced by similar 
fractional amounts.
Note that the improvement by 21\% results from the properties of Compton scatterings.
The improvement is somewhat modest, but is consistent with expectations based on 
Eqs.\ (\ref{b}) and (\ref{MDP}): for events with a certain modulation factor $\mu$ 
the MDP scales inversely proportional to $\mu$ (Equ.\ \ref{MDP}). For the 
``best events'' the modulation factor is close to 1; for energies $\ll\,m_{\rm e} c^2$ 
the average  modulation factor is 0.5. The decrease of the MDP should thus have 
a value of between 1 and 1/$\mu\,=$ 2.

For a real-world Compton polarimeter, a Maximum Likelihood analysis could be performed
with the Likelihood function given by Eqs.\ (\ref{M1}) and (\ref{M2}).
For experiments with non-uniform acceptance (e.g.\ for the 
ASTRO-H soft gamma-ray telescope of the ASTRO-H mission), 
the PDF of Equ.\ (\ref{M2}) should be replaced with a PDF obtained from Monte Carlo 
simulations. The PDF could then depend on a variety of measured parameters, e.g.\ 
the estimated energy of the primary event, the location of the first interaction, 
and the polar and azimuthal scattering angles. 
 
A Maximum Likelihood approach could also be used for the analysis 
of the data from photo-effect polarimeters as the gas polarimeters 
described in \citep{Cost:01,Soff:01,Depa:06,Mule:08,Mule:10,Blac:10}. 
The approach can be implemented by determining the modulation factor 
as a function of a set $\cal{O}_{\rm i}$ of suitable observables for 
each event, and by using the modulation factor $\mu({\cal{O}}_{\rm i})$ in Equ.\ (\ref{M2}).
One obvious choice is to use the reconstructed energy of the events: 
${\cal{O}}_{\rm i}\,=$ $\left\{E_i\right\}$.
Soffitta et al.\ \citep{Soff:01} and Depaola \& Longo \cite{Depa:06} 
discuss the pronounced energy dependence of the modulation factor. 
Some of the energy dependence stems from the atomic physics of the photon-absorption, 
i.e.\ the modulation factor for {\it s} electrons is larger than for {\it p} and {\it d} 
electrons and Auger electrons are emitted isotropically.
Some of the energy dependence stems from the detection technique as the 
properties of the recorded photoelectron tracks depend on the photon energy. 
One may use additional properties of an event which have an impact on
the effective modulation factor, e.g., the track length or the number of 
triggered pixels, the ellipticity of the track image when analyzed 
with a second moment analysis, or the estimated error on the reconstructed 
azimuthal scattering angle. 
The usage of $\mu({\cal{O}}_{\rm i})$ in Equ.\ (\ref{M2}) will improve the
sensitivity of the polarimeter as the individual events are weighted 
according to their information content. \\[2ex]
{\large \bf Acknowledgements:} 
HK acknowledges NASA for support from the APRA program under the grant NNX10AJ56G 
and support from the high-energy physics division of the DOE. HK thanks 
Martin Israel and an anonymous referee for helpful comments. HK is grateful for
support by the Washington University McDonnell Center for the 
Space Sciences.

\end{document}